\documentclass[a4paper,11pt]{article}
\usepackage{pos}
\usepackage{amsmath}
\usepackage{amsfonts}
\usepackage{amssymb}
\usepackage{xspace}
\usepackage{newtxtext,newtxmath}
\usepackage{bm}
\usepackage[normalem]{ulem} 

\newcommand{\ucm}{Departamento de F\'isica Te\'orica and IPARCOS, Universidad Complutense de Madrid, 28040 Madrid, Spain}
\newcommand{\ific}{Instituto de F\'{\i}sica Corpuscular (centro mixto CSIC-UV),
Institutos de Investigaci\'on de Paterna,
C/Catedr\'atico Jos\'e Beltr\'an 2, E-46980 Paterna, Valencia, Spain}
\newcommand{\ugr}{Departamento de F\'isica At\'omica, Molecular y Nuclear and Instituto Carlos I de F\'isica Te\'orica y Computacional, Universidad de
Granada, E-18071, Granada, Spain}
\renewcommand*{\vec}[1]{\ensuremath{\bm{\mathrm{#1}}}}

\title{Description of femtoscopic correlations with realistic pion-kaon interactions: the $\kappa/K^*(700)$ case}

\author*[1]{A. Canoa}
\author[2]{M. Albaladejo}
\author[2]{J. Nieves}
\author[1]{J.R. Peláez}
\author[3]{E. Ruiz Arriola}
\author[1]{J. Ruiz de Elvira}

\affiliation[1]{\ucm}

\affiliation[2]{\ific}

\affiliation[3]{\ugr}

\emailAdd{acanoa@ucm.es}
\emailAdd{Miguel.Albaladejo@ific.uv.es}
\emailAdd{Juan.M.Nieves@ific.uv.es}
\emailAdd{jrpelaez@fis.ucm.es}
\emailAdd{earriola@ugr.es}
\emailAdd{jacobore@ucm.es}

\abstract{In this work, we show how $\pi^+ K_S$ femtoscopic correlations, recently reported by ALICE collaboration in $pp$ collisions, can be well described taking into account relativistic corrections and using realistic $\pi K$ interactions.  These are obtained from a dispersive analysis of scattering data, which provides an accurate and model-independent description of the $\kappa/K^*_0(700)$ resonance pole.  The chiral symmetry suppression of the $\pi K$ interactions at low energies and the non-ordinary features of the $\kappa/K^*_0(700)$ seem to suggest that the $\pi^+ K_S$ source radius might be surprisingly smaller than for other hadronic processes when using the standard and simple Lednicky-Lyuboshits factorization approximation.}

\FullConference{10th International Conference on Quarks and Nuclear Physics (QNP2024).\\
8-12 July, 2024 .\\
Barcelona, Spain.\\
~ \\
IPARCOS-UCM-25-005 
}

\begin{document}
\maketitle

\section{Introduction}
The study of femtoscopic correlations between particle pairs was originally devised to determine the space-time structure of a particle emitting source \cite{Lisa:2005dd}.  In recent times, however, this formalism has gained an increasing interest in relativistic collisions carried out at LHC.  Benefiting from the huge statistics available there, it is now also seen as a tool to extract information on how the pair interacts, mainly when the interaction is not accessible by other means.  That is the case of hyperon-nucleon reactions, where the scattering data are extremely scarce \cite{Fabbietti:2020bfg}. 
The femtoscopy method could also prove valuable for light mesons.  
ALICE collaboration at CERN suggests that the femtoscopy formalism could shed light on open questions in the low-energy region, such as the nature of promising exotic resonances. 
 Examples are the $a_0 (980)$~\cite{ALICE:2021ovd} and our case of interest, the $\kappa/K^*_0(700)$, discussed in~\cite{ALICE:2023eyl} using $\pi^{\pm} K_S$ femtoscopic correlations.  In this talk, we identify several caveats and propose the corresponding improvements on the simple but rather standard theoretical formalism used in~\cite{ALICE:2023eyl}.

Experimentally, the correlation function is measured as the ratio between the momentum distribution of particle pairs in the same event and a reference momentum distribution of pairs from different events. This ratio approaches unity when the momenta are large and no correlations are present. The theoretical correlation function $C(k^*)$, with $\vec{k^*}$ the pair center-of-mass momentum (CM), is typically described using the Koonin-Pratt formalism~\cite{Koonin:1977fh,Pratt:1984su},~$C(k^*)=\int S(r) |\psi(\vec{k^*},\vec{r})|^2 d^3 \vec{r}$, as the spatial overlap of the squared modulus of the pair wave-function $\psi$  and the source $S$. The source is usually modeled as a normalized Gaussian distribution, depending on the relative distance between the particle pair $\vec{r}$ and the source radius $R$.
\section{ALICE Data and model}
In \cite{ALICE:2023eyl}, the ALICE collaboration reports remarkable data from $\pi^{+}K_S$ femtoscopic s-wave correlations. 
 The data are collected in 3 sets with different cuts, according to the multiplicity classes and the dependence of transverse momentum of the measured pairs.  In order to extract the actual correlations, the raw correlation data have been thoroughly treated to remove non-femtoscopic effects, such as soft parton fragmentation, and the contribution of the K$^* (892) $, at the cost of introducing two more fit parameters.  ALICE describes the remaining s-wave correlations with the widely used and popular model of Lednicky and Lyuboshits (LL)~\cite{Lednicky:1981su}:
\begin{equation}
C_{LL}(k^*)=1+\frac{\lambda \alpha}{2}\left[\left|\frac{f(k^*)}{R} \right|^2+\frac{4 \text{Re}(f(k^*))}{\sqrt{\pi}R}F_1(2k^* R)-2\frac{\text{Im}(f(k^*))}{R}F_2(2k^*R) +\Delta C\right].
\end{equation}
The model is non-relativistic (NR), it assumes a Gaussian form of the source and the on-shell factorization of the scattering amplitude $f(k^*)$.  $F_1$ and $F_2$ are two known analytic functions that appear when integrating the intermediate pion and kaon propagators with total momentum $k^*$ together with the source.  Then, $\alpha$  is a symmetry factor $(1/2$ in our case), $R$ is the source radius, and $\lambda$, called correlation strength, accounts for the purity of the genuine pairs. 
 Both $R$ and $\lambda$ are free parameters.  The $\Delta C$ term is a correction to the LL formula, which assumes, by construction, outgoing free spherical waves instead of the true scattered waves.

In addition, for $f(k^*)$, ALICE assumes that the $\pi K$ s-wave scattering is purely elastic with isospin $I=1/2$.  They also impose that the $I=1/2$ wave is completely dominated by the $\kappa/K^*_0(700)$ resonance, which they parametrize as a relativistic Breit-Wigner (BW) amplitude,
\begin{equation}
f(k^*)=\frac{\gamma}{M_R^2-s-i \gamma k^*},
\end{equation}
where $M_R$ and $\gamma$ are fit parameters that represent the mass and width of the resonance respectively, while $s$ denotes the total relativistic energy in the CM frame.

Now that the ALICE model is presented, some concerns arise. First, there is no $I=3/2$  contribution, so it neglects any inelastic effect. In nature, $\pi^+K_S$ couples to $\pi^+ K_L$ or $\pi^0 K^+$, which are different combinations of isospin $1/2$ and $3/2$.  Second, the pion mass $M_\pi\approx 0.14$ GeV, is significantly smaller than the maximum momentum considered, $k^*\approx 0.76$ GeV, pointing to the need for relativistic corrections. Regarding the choice of the BW to model the final state interaction (FSI), the very broad and not ordinary $\kappa/K^*_0(700)$ resonance does not saturate unitarity, unlike a typical BW-like pole. Besides, an ordinary BW overestimates the interaction near threshold, 
since, in real life, it is suppressed by the SU(3) chiral symmetry-breaking dynamics that dominate in this region. Indeed, Fig.~\ref{fig:BWversusdata} shows how the BW (red curve) is unable to reproduce the $\pi K$ scattering data (black square and circle points)~\cite{Aston:1987ir,Estabrooks:1977xe}. Moreover, the fit values of $M_R$ and $\gamma$ in~\cite{ALICE:2023eyl} are inconsistent for each dataset, although they should describe the same final state $\pi K$ interaction.

\section{Improvements}
\begin{figure}[ht!]
\centering
\includegraphics[width=0.6\textwidth]{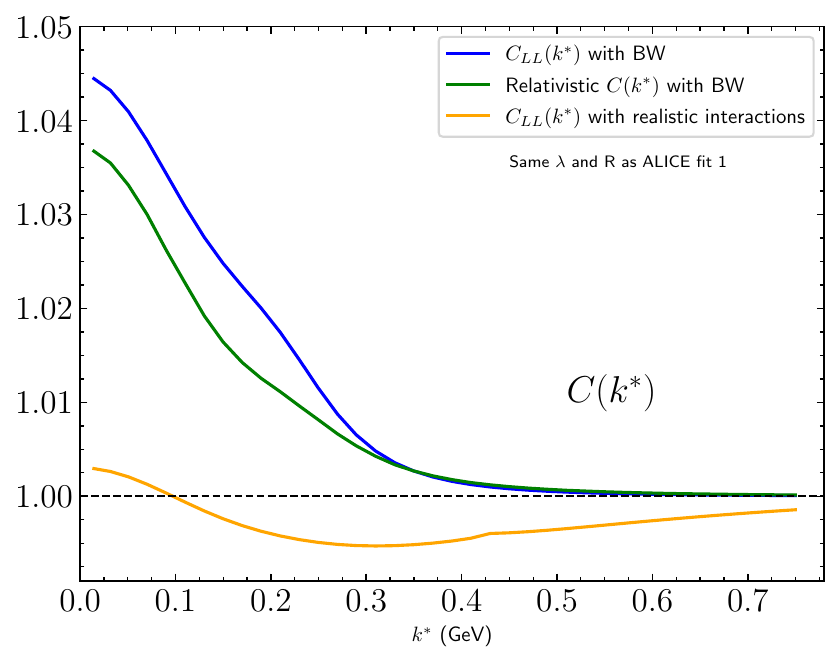}
 \caption{\rm \label{fig:comparison} 
 The blue curve is the result of ALICE (fit 1, Fig.~2 in~\cite{ALICE:2023eyl})
 using the LL formula with an elastic BW for the $K_S \pi^\pm$ interaction. The green line just replaces the LL non-relativistic formula with
 our relativistic equation. The orange line keeps the LL formulation but replaces the BW with a realistic description of $K\pi$ scattering data, including its $I=3/2$ component, and also accounts for the effect of coupled channels.
}
\end{figure}

In this work, we will improve each of the above points.  In order to see their individual effects, we will first implement them separately and only consider them together at the end.  Thus, first, we keep the FSI unchanged and implement relativistic corrections, constructing the wave-function of the pair from the Bethe-Salpeter equation~\cite{PhysRev.84.1232}.  This essentially consists of replacing the usual NR propagators and integral measure with their relativistic counterparts~\cite{RuizdeElvira:2018hsv,Albaladejo:2024lam}.  Surprisingly, the correction is already large ($\approx 20 \%)$ at the threshold due to the virtual momenta integration needed to obtain the relativistic versions of $F_1$ and $F_2$.  In Fig.~\ref{fig:comparison}, we plot $C(k^*)$, where the effect of relativistic corrections is seen to be important also in absolute terms.  At high energies, the relativistic (green) and NR (blue) curves look identical because the correlations become very small anyway.

\begin{figure}[ht!]
\centering
\includegraphics[width=0.6
\textwidth]{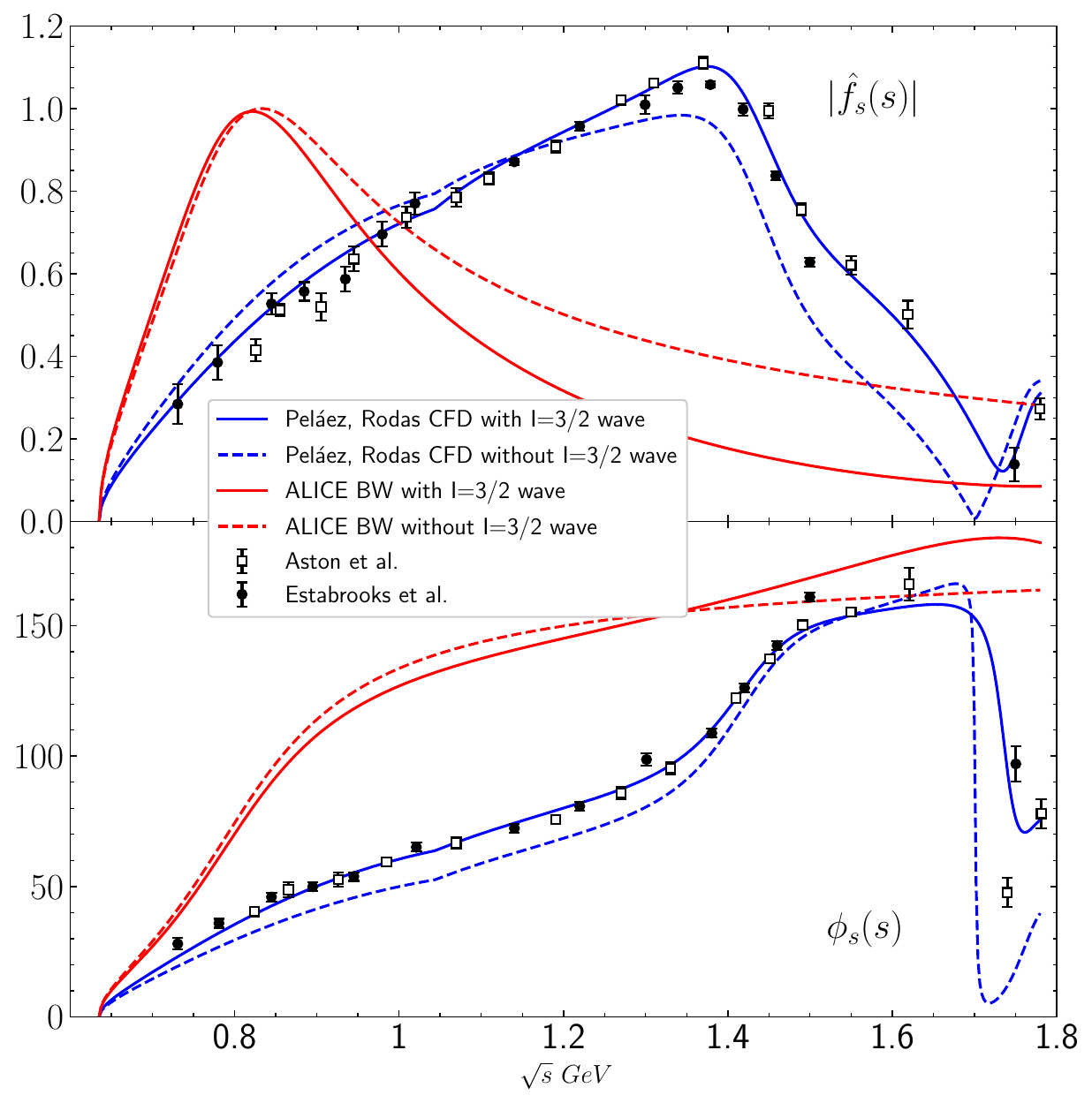}
\caption{\rm \label{fig:BWversusdata} $K \pi$ scattering data
 in the scalar channel $K^+ \pi^-$ (Aston~\cite{Aston:1987ir} in open square markers, Estabrooks~\cite{Estabrooks:1977xe} in filled circular markers).  The scalar channel amplitude is defined as $\hat f_S\equiv \hat f_0^{1/2}+\hat f_0^{3/2}/2=\vert \hat f_S\vert \exp{(i\phi_S)}$. Top: $\vert \hat f_S\vert$, Bottom: $\phi_S$. The red line is the BW used by ALICE (Fit 1. The other two are similar). Blue lines correspond to the dispersive analysis of \cite{Pelaez:2020gnd}. Dashed lines are obtained ignoring the $I=3/2$ contribution.}
\end{figure}

Regarding the interaction, we now keep the NR LL formula but employ a dispersive parametrization of the $\pi K$ scattering data, the so-called Constrained Fits to Data (CFD)~\cite{Pelaez:2020gnd}, instead of the elastic BW proposed by ALICE to model the FSI.  As shown in Fig.~\ref{fig:BWversusdata} (blue curve), this provides an accurate description of the $\pi K$ scattering data, extracted from~\cite{Aston:1987ir,Estabrooks:1977xe} in the scalar combination $\pi^+ K^-$. It is worth noting that neglecting the isospin $I = \frac{3}{2}$ component (red and blue dashed lines) has a small but noticeable effect in the scattering data.  Moreover, for the $\pi^+ K_S$ channel we are interested in for femtoscopy, the $I=3/2$ component is enhanced by an additional factor of 4.  Additionally, an inelastic term is needed in the correlation function to consider all channels coupled to $\pi^+ K_S$.  Fig.~\ref{fig:comparison} shows that $C(k^*)$ significantly decreases for the same $R$ and $\lambda$ once the CFD parametrization and coupled channels are included (orange curve), and this effect is even more pronounced than the relativistic corrections mentioned earlier.

\section{Preliminary results}
\begin{figure}[ht!]
\centering
\includegraphics[width=0.55\textwidth]{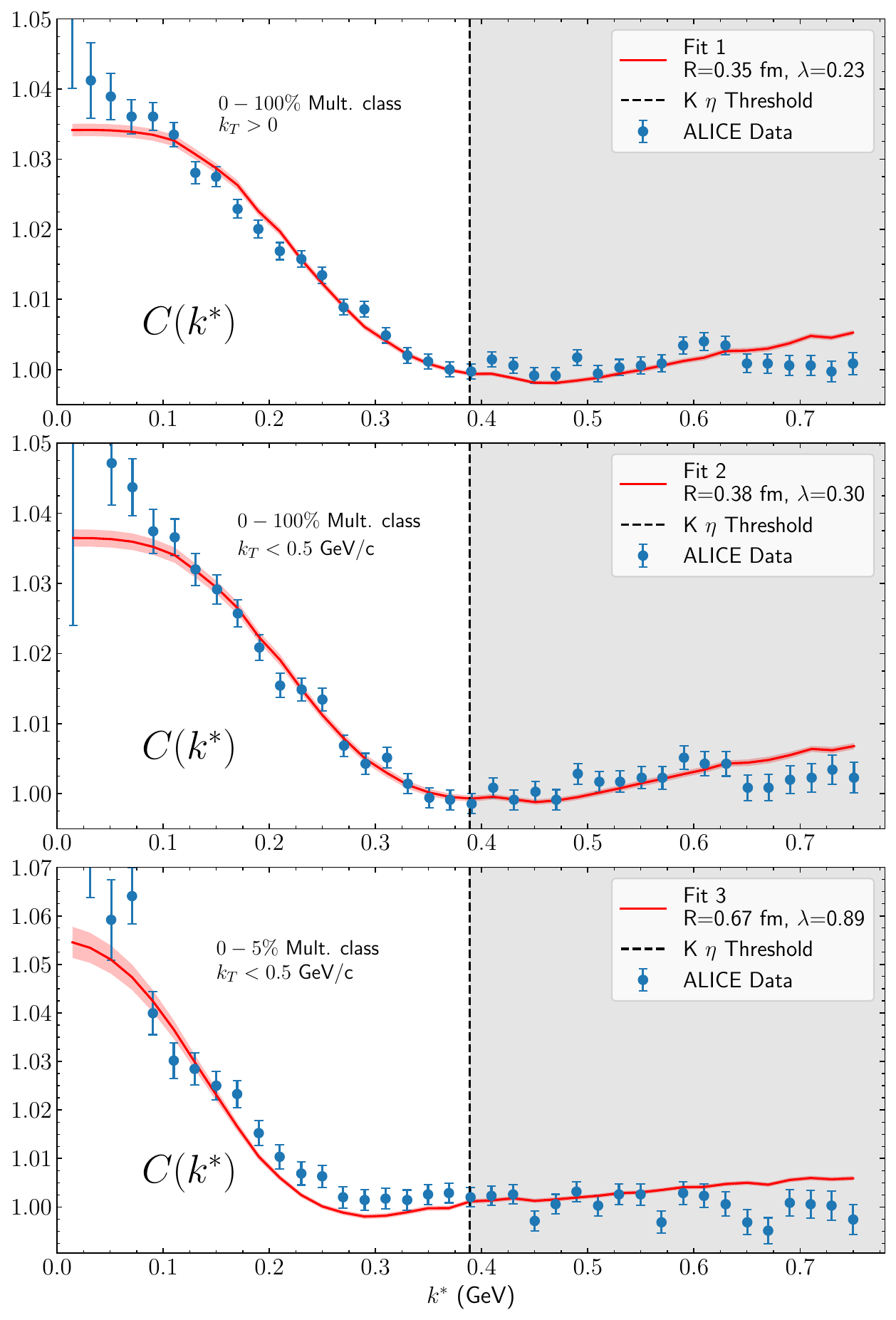}
 \caption{\rm \label{fig:ourresults} 
Our preliminary description of femtoscopic s-wave correlations for the 3 ALICE datasets \cite{ALICE:2023eyl}. The red lines and bands are the results of our fits using our formalism with realistic $K\pi$ interactions and relativistic corrections.  The dashed black line marks the opening of the $K \eta$ threshold, where our formalism is formally no longer valid, although it still yields a fairly good qualitative description.
}
\end{figure}

When we implement simultaneously relativistic corrections, coupled channels, and the dispersive parametrization discussed above, we obtain the fits to the data shown in Fig.~\ref{fig:ourresults}.  In contrast to the ALICE fits, which required 6 parameters for each dataset, our approach only needs 3 fit parameters, namely, an overall normalization factor $\kappa$ (which differs from $1$ by less than $1 \%$), the source radius $R$, and the correlation strength $\lambda$.  The reason for this reduction is that we adopt the same $K \pi$ interaction for the 3 sets, fixed from scattering data, whereas ALICE fits $M_R, \gamma$  and an extra parameter used to
obtain the scalar femtoscopic correlations from
their raw correlations.  Our curves reproduce nicely the correlation data, particularly in the elastic region below the $K \eta$ threshold. Note that in Fig.~\ref{fig:ourresults} the values that we obtain for the source radius are smaller than those typical in the literature (around 1 fm), which may raise questions about the validity of the widely used on-shell factorization models for relativistic light mesons, as well as the conclusions drawn about the nature of the $\kappa/K^*_0(700)$ in~\cite{ALICE:2023eyl}.

\acknowledgments
A. C. thanks the organizers of  QNP2024 for the opportunity to participate in his first conference. Work supported by the Spanish Ministerio de Ciencia e Innovación (MICINN) and European FEDER funds under Contracts PID2020-112777GB-I00, PID2022-136510NB-C31, PID2023-147072NB-I00, PID2023-147458NB-C21 and CEX2023-001292S; by Generalitat Valenciana (GVA) under contracts PROMETEO/2020/023 and CIPROM/2023/59, by Junta de Andalucía grant FQM-225 and European Union’s Horizon 2020 research and innovation program (grant~824093). A. C. is supported by Comunidad de Madrid predoctoral grant PIPF-2023 (Order~5027/2023); J. R. E. by the Ram\'on y Cajal program (RYC2019-027605-I) of MINECO, and M. A. by GVA Grant~CIDEGENT/2020/002 and MICINN Ramón y Cajal program Grant~RYC2022-038524-I.    

\begin{thebibliography}{13}%
\makeatletter
\providecommand \@ifxundefined [1]{%
 \@ifx{#1\undefined}
}%
\providecommand \@ifnum [1]{%
 \ifnum #1\expandafter \@firstoftwo
 \else \expandafter \@secondoftwo
 \fi
}%
\providecommand \@ifx [1]{%
 \ifx #1\expandafter \@firstoftwo
 \else \expandafter \@secondoftwo
 \fi
}%
\providecommand \natexlab [1]{#1}%
\providecommand \enquote  [1]{``#1''}%
\providecommand \bibnamefont  [1]{#1}%
\providecommand \bibfnamefont [1]{#1}%
\providecommand \citenamefont [1]{#1}%
\providecommand \href@noop [0]{\@secondoftwo}%
\providecommand \href [0]{\begingroup \@sanitize@url \@href}%
\providecommand \@href[1]{\@@startlink{#1}\@@href}%
\providecommand \@@href[1]{\endgroup#1\@@endlink}%
\providecommand \@sanitize@url [0]{\catcode `\\12\catcode `\$12\catcode `\&12\catcode `\#12\catcode `\^12\catcode `\_12\catcode `\%12\relax}%
\providecommand \@@startlink[1]{}%
\providecommand \@@endlink[0]{}%
\providecommand \url  [0]{\begingroup\@sanitize@url \@url }%
\providecommand \@url [1]{\endgroup\@href {#1}{\urlprefix }}%
\providecommand \urlprefix  [0]{URL }%
\providecommand \Eprint [0]{\href }%
\providecommand \doibase [0]{http://dx.doi.org/}%
\providecommand \selectlanguage [0]{\@gobble}%
\providecommand \bibinfo  [0]{\@secondoftwo}%
\providecommand \bibfield  [0]{\@secondoftwo}%
\providecommand \translation [1]{[#1]}%
\providecommand \BibitemOpen [0]{}%
\providecommand \bibitemStop [0]{}%
\providecommand \bibitemNoStop [0]{.\EOS\space}%
\providecommand \EOS [0]{\spacefactor3000\relax}%
\providecommand \BibitemShut  [1]{\csname bibitem#1\endcsname}%
\let\auto@bib@innerbib\@empty
\bibitem [{\citenamefont {Lisa}\ \emph {et~al.}(2005)\citenamefont {Lisa}, \citenamefont {Pratt}, \citenamefont {Soltz},\ and\ \citenamefont {Wiedemann}}]{Lisa:2005dd}%
  \BibitemOpen
  \bibfield  {author} {\bibinfo {author} {\bibfnamefont {M.~A.}\ \bibnamefont {Lisa}}, \bibinfo {author} {\bibfnamefont {S.}~\bibnamefont {Pratt}}, \bibinfo {author} {\bibfnamefont {R.}~\bibnamefont {Soltz}}, \ and\ \bibinfo {author} {\bibfnamefont {U.}~\bibnamefont {Wiedemann}},\ }\href {\doibase 10.1146/annurev.nucl.55.090704.151533} {\bibfield  {journal} {\bibinfo  {journal} {Ann. Rev. Nucl. Part. Sci.}\ }\textbf {\bibinfo {volume} {55}},\ \bibinfo {pages} {357} (\bibinfo {year} {2005})},\ \Eprint {http://arxiv.org/abs/nucl-ex/0505014} {arXiv:nucl-ex/0505014} \BibitemShut {NoStop}%
\bibitem [{\citenamefont {Fabbietti}\ \emph {et~al.}(2021)\citenamefont {Fabbietti}, \citenamefont {Mantovani~Sarti},\ and\ \citenamefont {Vazquez~Doce}}]{Fabbietti:2020bfg}%
  \BibitemOpen
  \bibfield  {author} {\bibinfo {author} {\bibfnamefont {L.}~\bibnamefont {Fabbietti}}, \bibinfo {author} {\bibfnamefont {V.}~\bibnamefont {Mantovani~Sarti}}, \ and\ \bibinfo {author} {\bibfnamefont {O.}~\bibnamefont {Vazquez~Doce}},\ }\href {\doibase 10.1146/annurev-nucl-102419-034438} {\bibfield  {journal} {\bibinfo  {journal} {Ann. Rev. Nucl. Part. Sci.}\ }\textbf {\bibinfo {volume} {71}},\ \bibinfo {pages} {377} (\bibinfo {year} {2021})},\ \Eprint {http://arxiv.org/abs/2012.09806} {arXiv:2012.09806 [nucl-ex]} \BibitemShut {NoStop}%
\bibitem [{\citenamefont {Acharya}\ \emph {et~al.}(2022)\citenamefont {Acharya} \emph {et~al.}}]{ALICE:2021ovd}%
  \BibitemOpen
  \bibfield  {author} {\bibinfo {author} {\bibfnamefont {S.}~\bibnamefont {Acharya}} \emph {et~al.} (\bibinfo {collaboration} {ALICE}),\ }\href {\doibase 10.1016/j.physletb.2022.137335} {\bibfield  {journal} {\bibinfo  {journal} {Phys. Lett. B}\ }\textbf {\bibinfo {volume} {833}},\ \bibinfo {pages} {137335} (\bibinfo {year} {2022})},\ \Eprint {http://arxiv.org/abs/2111.06611} {arXiv:2111.06611 [nucl-ex]} \BibitemShut {NoStop}%
\bibitem [{\citenamefont {Acharya}\ \emph {et~al.}(2024)\citenamefont {Acharya} \emph {et~al.}}]{ALICE:2023eyl}%
  \BibitemOpen
  \bibfield  {author} {\bibinfo {author} {\bibfnamefont {S.}~\bibnamefont {Acharya}} \emph {et~al.} (\bibinfo {collaboration} {ALICE}),\ }\href {\doibase 10.1016/j.physletb.2024.138915} {\bibfield  {journal} {\bibinfo  {journal} {Phys. Lett. B}\ }\textbf {\bibinfo {volume} {856}},\ \bibinfo {pages} {138915} (\bibinfo {year} {2024})},\ \Eprint {http://arxiv.org/abs/2312.12830} {arXiv:2312.12830 [hep-ex]} \BibitemShut {NoStop}%
\bibitem [{\citenamefont {Koonin}(1977)}]{Koonin:1977fh}%
  \BibitemOpen
  \bibfield  {author} {\bibinfo {author} {\bibfnamefont {S.~E.}\ \bibnamefont {Koonin}},\ }\href {\doibase 10.1016/0370-2693(77)90340-9} {\bibfield  {journal} {\bibinfo  {journal} {Phys. Lett. B}\ }\textbf {\bibinfo {volume} {70}},\ \bibinfo {pages} {43} (\bibinfo {year} {1977})}\BibitemShut {NoStop}%
\bibitem [{\citenamefont {Pratt}(1984)}]{Pratt:1984su}%
  \BibitemOpen
  \bibfield  {author} {\bibinfo {author} {\bibfnamefont {S.}~\bibnamefont {Pratt}},\ }\href {\doibase 10.1103/PhysRevLett.53.1219} {\bibfield  {journal} {\bibinfo  {journal} {Phys. Rev. Lett.}\ }\textbf {\bibinfo {volume} {53}},\ \bibinfo {pages} {1219} (\bibinfo {year} {1984})}\BibitemShut {NoStop}%
\bibitem [{\citenamefont {Lednicky}\ and\ \citenamefont {Lyuboshits}(1981)}]{Lednicky:1981su}%
  \BibitemOpen
  \bibfield  {author} {\bibinfo {author} {\bibfnamefont {R.}~\bibnamefont {Lednicky}}\ and\ \bibinfo {author} {\bibfnamefont {V.~L.}\ \bibnamefont {Lyuboshits}},\ }\href@noop {} {\bibfield  {journal} {\bibinfo  {journal} {Yad. Fiz.}\ }\textbf {\bibinfo {volume} {35}},\ \bibinfo {pages} {1316} (\bibinfo {year} {1981})}\BibitemShut {NoStop}%
\bibitem [{\citenamefont {Aston}\ \emph {et~al.}(1988)\citenamefont {Aston} \emph {et~al.}}]{Aston:1987ir}%
  \BibitemOpen
  \bibfield  {author} {\bibinfo {author} {\bibfnamefont {D.}~\bibnamefont {Aston}} \emph {et~al.},\ }\href {\doibase 10.1016/0550-3213(88)90028-4} {\bibfield  {journal} {\bibinfo  {journal} {Nucl. Phys.}\ }\textbf {\bibinfo {volume} {B296}},\ \bibinfo {pages} {493} (\bibinfo {year} {1988})}\BibitemShut {NoStop}%
\bibitem [{\citenamefont {Estabrooks}\ \emph {et~al.}(1978)\citenamefont {Estabrooks}, \citenamefont {Carnegie}, \citenamefont {Martin}, \citenamefont {Dunwoodie}, \citenamefont {Lasinski},\ and\ \citenamefont {Leith}}]{Estabrooks:1977xe}%
  \BibitemOpen
  \bibfield  {author} {\bibinfo {author} {\bibfnamefont {P.}~\bibnamefont {Estabrooks}}, \bibinfo {author} {\bibfnamefont {R.~K.}\ \bibnamefont {Carnegie}}, \bibinfo {author} {\bibfnamefont {A.~D.}\ \bibnamefont {Martin}}, \bibinfo {author} {\bibfnamefont {W.~M.}\ \bibnamefont {Dunwoodie}}, \bibinfo {author} {\bibfnamefont {T.~A.}\ \bibnamefont {Lasinski}}, \ and\ \bibinfo {author} {\bibfnamefont {D.~W. G.~S.}\ \bibnamefont {Leith}},\ }\href {\doibase 10.1016/0550-3213(78)90238-9} {\bibfield  {journal} {\bibinfo  {journal} {Nucl. Phys.}\ }\textbf {\bibinfo {volume} {B133}},\ \bibinfo {pages} {490} (\bibinfo {year} {1978})}\BibitemShut {NoStop}%
\bibitem [{\citenamefont {Salpeter}\ and\ \citenamefont {Bethe}(1951)}]{PhysRev.84.1232}%
  \BibitemOpen
  \bibfield  {author} {\bibinfo {author} {\bibfnamefont {E.~E.}\ \bibnamefont {Salpeter}}\ and\ \bibinfo {author} {\bibfnamefont {H.~A.}\ \bibnamefont {Bethe}},\ }\href {\doibase 10.1103/PhysRev.84.1232} {\bibfield  {journal} {\bibinfo  {journal} {Phys. Rev.}\ }\textbf {\bibinfo {volume} {84}},\ \bibinfo {pages} {1232} (\bibinfo {year} {1951})}\BibitemShut {NoStop}%
\bibitem [{\citenamefont {Ruiz~de Elvira}\ and\ \citenamefont {Ruiz~Arriola}(2018)}]{RuizdeElvira:2018hsv}%
  \BibitemOpen
  \bibfield  {author} {\bibinfo {author} {\bibfnamefont {J.}~\bibnamefont {Ruiz~de Elvira}}\ and\ \bibinfo {author} {\bibfnamefont {E.}~\bibnamefont {Ruiz~Arriola}},\ }\href {\doibase 10.1140/epjc/s10052-018-6342-7} {\bibfield  {journal} {\bibinfo  {journal} {Eur. Phys. J. C}\ }\textbf {\bibinfo {volume} {78}},\ \bibinfo {pages} {878} (\bibinfo {year} {2018})},\ \Eprint {http://arxiv.org/abs/1807.10837} {arXiv:1807.10837 [hep-ph]} \BibitemShut {NoStop}%
\bibitem [{\citenamefont {Albaladejo}\ \emph {et~al.}(2024)\citenamefont {Albaladejo}, \citenamefont {Feijoo}, \citenamefont {Nieves}, \citenamefont {Oset},\ and\ \citenamefont {Vida\~na}}]{Albaladejo:2024lam}%
  \BibitemOpen
  \bibfield  {author} {\bibinfo {author} {\bibfnamefont {M.}~\bibnamefont {Albaladejo}}, \bibinfo {author} {\bibfnamefont {A.}~\bibnamefont {Feijoo}}, \bibinfo {author} {\bibfnamefont {J.}~\bibnamefont {Nieves}}, \bibinfo {author} {\bibfnamefont {E.}~\bibnamefont {Oset}}, \ and\ \bibinfo {author} {\bibfnamefont {I.}~\bibnamefont {Vida\~na}},\ }\href@noop {} {\  (\bibinfo {year} {2024})},\ \Eprint {http://arxiv.org/abs/2410.08880} {arXiv:2410.08880 [hep-ph]} \BibitemShut {NoStop}%
\bibitem [{\citenamefont {Pel\'aez}\ and\ \citenamefont {Rodas}(2022)}]{Pelaez:2020gnd}%
  \BibitemOpen
  \bibfield  {author} {\bibinfo {author} {\bibfnamefont {J.~R.}\ \bibnamefont {Pel\'aez}}\ and\ \bibinfo {author} {\bibfnamefont {A.}~\bibnamefont {Rodas}},\ }\href {\doibase 10.1016/j.physrep.2022.03.004} {\bibfield  {journal} {\bibinfo  {journal} {Phys. Rept.}\ }\textbf {\bibinfo {volume} {969}},\ \bibinfo {pages} {1} (\bibinfo {year} {2022})},\ \Eprint {http://arxiv.org/abs/2010.11222} {arXiv:2010.11222 [hep-ph]} \BibitemShut {NoStop}%
\end{thebibliography}
%

\end{document}